\documentstyle[emulateapj,psfig,apjfonts]{article}

\def\etal{{\rm et~al.\ }}

\def\HII{H\,{\sc ii}}
\def\kms{km~s$^{-1}$}

\received{}
\accepted{}
\journalid{}{}
\articleid{}{}

\slugcomment{}

\lefthead{CRAWFORD ET AL}
\righthead{A RADIO SUPERNOVA REMNANT ASSOCIATED WITH PSR~J1119--6127}

\begin{document}

\title{A Radio Supernova Remnant Associated with the  
Young Pulsar J1119--6127}

\author{Fronefield Crawford\altaffilmark{1,8},  
B. M. Gaensler\altaffilmark{1,6},  
V. M. Kaspi\altaffilmark{1,2,7}, \\
R. N. Manchester\altaffilmark{3}, 
F. Camilo\altaffilmark{4}, 
A. G. Lyne\altaffilmark{5}
and M. J. Pivovaroff\altaffilmark{1,9}} 

\altaffiltext{1}{Department of Physics and Center for Space Research,
Massachusetts Institute of Technology, 70 Vassar Street,
Cambridge, MA 02139}

\altaffiltext{2}{Department of Physics, Rutherford Physics Building, 
McGill University, 3600 University Street, Montreal, Quebec, H3A 2T8, Canada}

\altaffiltext{3}{Australia Telescope National Facility, CSIRO,
PO Box 76, Epping, NSW 1718, Australia}

\altaffiltext{4}{Columbia Astrophysics Laboratory, Columbia
University, 550 W. 120th Street, New York, NY 10027}

\altaffiltext{5}{University of Manchester, Jodrell Bank Observatory,
Macclesfield, Cheshire SK11 9DL, United Kingdom}

\altaffiltext{6}{Hubble Fellow}

\altaffiltext{7}{Alfred P. Sloan Research Fellow}

\altaffiltext{8}{Current address: Management and Data Systems Division,
Lockheed Martin Corporation, PO Box 8048, Philadelphia, PA 19101}

\altaffiltext{9}{Current address: Therma-Wave Inc., 1250 Reliance Way,
Fremont, CA 94539}

\begin{abstract}

We report on Australia Telescope Compact Array observations in the
direction of the young high magnetic-field pulsar J1119--6127. In
the resulting images we identify a non-thermal radio shell
of diameter $15'$, 
which we classify as a previously uncatalogued
young supernova remnant, G292.2--0.5.  This
supernova remnant is positionally coincident with PSR~J1119--6127, and
we conclude that the two objects are physically associated. No radio
emission is detected from any pulsar wind nebula (PWN) associated with
the pulsar; our observed upper limits are consistent with the
expectation that high magnetic-field pulsars produce radio nebulae
which fade rapidly. This system suggests a possible explanation for the
lack of an associated radio pulsar and/or PWN in many supernova remnants.

\end{abstract}


\section{Introduction} 

%\subsection{Scientific Motivation}

Important insight into the formation and evolution
of pulsars and supernova remnants (SNRs) can be obtained
by finding physical associations between the two
classes of object. Pulsars are believed
to be formed in supernovae which result from the collapse
of massive stars, a scenario which implies
that young pulsars and SNRs should be associated.
However, there are only about a dozen cases
in which such a pairing can be seriously claimed
(\cite{kas98}). 
It is still a matter of debate as to why there are so few
pulsar / SNR associations.  While there is no doubt that selection
effects associated with detecting each type of object can at least
partly account for the deficit in associations (\cite{gj95c}), 
there is mounting
evidence that a significant fraction of SNRs may be associated with
objects with very different properties from traditional radio pulsars
(\cite{gv00}).  Establishing an association between a SNR and any
sort of compact object addresses this fundamental issue of what is left
behind when a star explodes.  
In the particular case of associations of radio pulsars with SNRs,
studies of such systems provide important
constraints on pulsar properties such as their initial spin-periods,
magnetic fields, radio luminosities, birth-rates and
beaming fractions (e.g.\ \cite{bj98}). 

Meanwhile, the young pulsars usually associated 
with SNRs lose their rotational kinetic energy at a rapid
rate. These pulsars deposit this energy into their surroundings
in the form of a magnetized relativistic particle
wind. This wind can interact with the ambient medium to 
produce an observable
pulsar wind nebula (PWN), the study of which can 
provide information on both the pulsar's wind and
surrounding environment (e.g.\ \cite{fggd96}).

\subsection{The young pulsar J1119--6127}

The 408-ms pulsar J1119--6127 (\cite{ckl+00})
was recently discovered in the Parkes multibeam pulsar survey
(\cite{lcm+00}; \cite{clm+00}). This pulsar has a very small
characteristic age $\tau_c \equiv P/2\dot{P} = 1.6$~kyr
and a very high dipole magnetic field, $B \approx 3.2\times10^{19}
(P\dot{P})^{1/2}$~G~$ = 4.1 \times 10^{13}$~G,
where 
$P$ and $\dot{P}$ are the pulsar's
spin-period and period derivative, respectively.
Assuming that neither the pulsar's magnetic moment nor moment of inertia
are evolving with time, it can be shown that an upper limit
on the pulsar's age is $1.7\pm0.1$~kyr (\cite{ckl+00}).
In Table~\ref{chap6:tab1} we compare the properties
of PSR~J1119--6127 to the other five known pulsars
for which $\tau_c < 5$~kyr. It can be seen that
each of these other young pulsars is associated with 
radio emission from a SNR and/or PWN, and it
thus seems reasonable to suppose that
PSR~J1119--6127 might similarly have an associated
SNR or PWN. 

\begin{figure}
%\plotone{figures/fig1.ps}
\centerline{\psfig{file=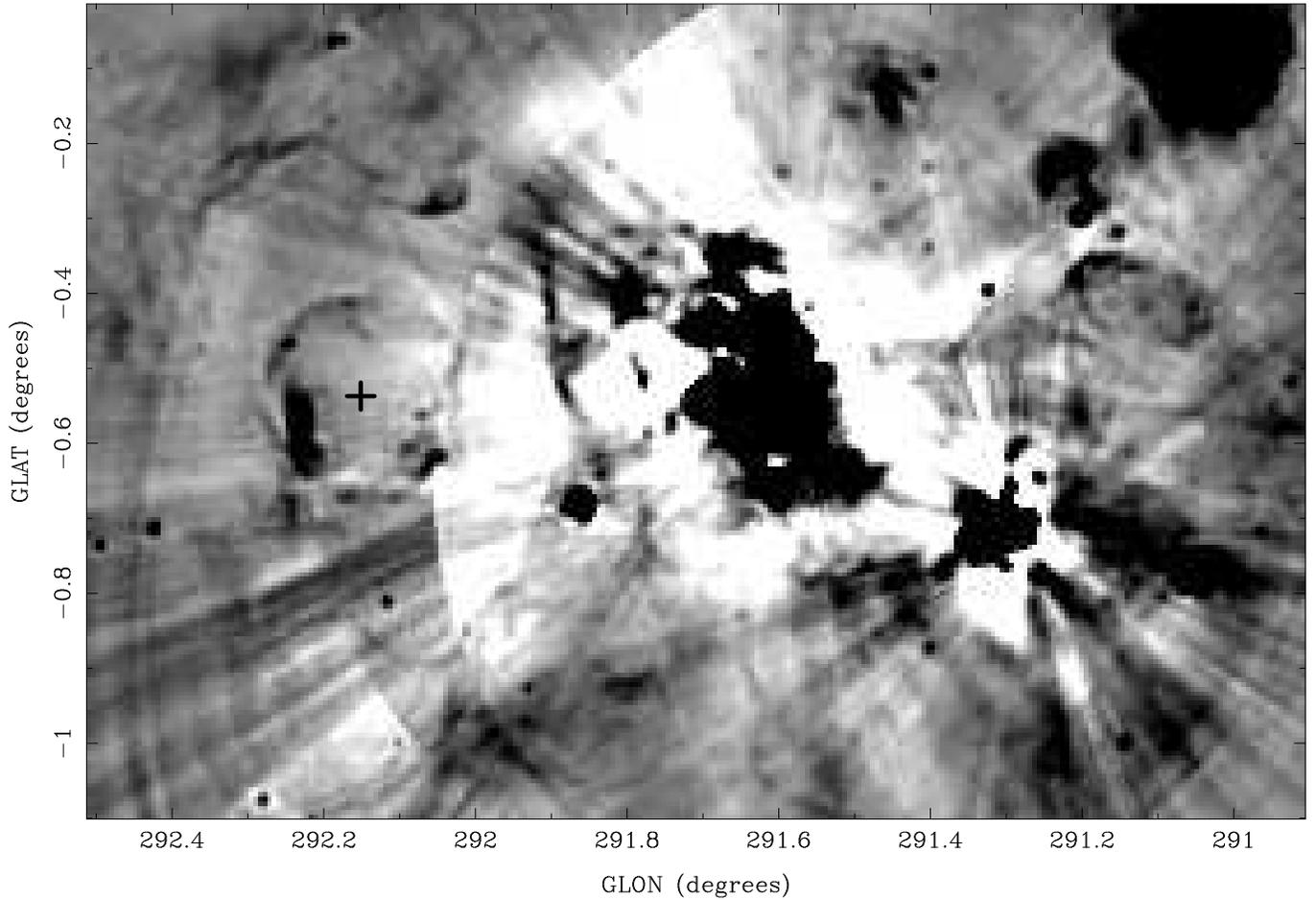,width=18cm,angle=270}}
\caption{An image 
of the region surrounding PSR~J1119$-$6127
taken from the Molonglo Galactic Plane Survey (MGPS; \cite{gcl98}),
at a frequency of 843~MHz and an angular resolution of $\sim1'$.
Here and in subsequent images, the
position of PSR~J1119--6127, as determined from pulsar-gating,
is indicated by a cross.  The bright \HII\ regions
NGC~3603 (G291.58--0.41)
and NGC~3576 (G291.28--0.71) can be seen to the west of the pulsar.}
\label{chap6:fig1}
\end{figure}

Indeed, examination of the highest-resolution radio data available for
the region (Fig.~\ref{chap6:fig1}; see also \cite{gcl98}) shows emission
in the vicinity of the pulsar suggestive of the morphology expected for
a SNR.  However, artifacts produced by the nearby bright  \HII\ regions
NGC~3603 and NGC~3576 (see \cite{dng99} and references
therein), along with further image defects occurring at
the boundary between adjacent fields, seriously limit the sensitivity
and image quality in this region.  We therefore have undertaken more
detailed observations of the field surrounding PSR~J1119--6127 in an
attempt to ascertain the nature of emission in the region, as described
in Section~\ref{sec_obs}.  In Section~\ref{sec_results} we show that
the emission seen in the pulsar's vicinity is a non-thermal shell,
G292.2--0.5, and set upper limits on the radio emission from a PWN at
the pulsar's position.  In Section~\ref{sec_assoc} we argue that
G292.2--0.5 is a previously unidentified SNR associated with
PSR~J1119--6127, while in Section~\ref{sec_pwn} we account for the
absence of a radio PWN as being a result of the pulsar's high magnetic field.
The results of X-ray observations of this system are presented in a
companion paper by Pivovaroff \etal (2000\nocite{pkc+00}).

% INCLUDE IN SECTION OUTLINE
%In addition, pulsar-gated observations allow the pulsed flux to be
%subtracted from the image, thereby giving a precise interferometric
%position for the pulsar and increased sensitivity to faint underlying
%emission which might otherwise be missed. 

% MOVE TO DISCUSSION
%Searching for a possible PWN
%associated with PSR J1119$-$6127 is particularly important since of
%the three youngest pulsars known, two (the Crab and PSR J0540$-$6919)
%have observed radio PWNe and one (PSR J1513$-$5908) does not (see
%Table \ref{chap6:tab1}). PSR J1513$-$5908 is similar to PSR
%J1119$-$6127 in two respects. Both of these pulsars have relatively
%long periods (150 ms for PSR J1513$-$5908 and 408 ms for PSR
%J1119$-$6127) compared to other young pulsars: the periods of the Crab
%pulsar and PSR J0540$-$6919 are 33 ms and 50 ms, respectively. Both
%PSRs J1513$-$5908 and J1119$-$6127 also have significantly larger
%magnetic field strengths than other young pulsars. The magnetic field
%is an important factor determining the evolution of both the pulsar
%and its PWN, as demonstrated below. As part of the imaging campaign to
%establish positions for young pulsars (\cite{cra00}), we determined an
%accurate position for PSR J1119$-$6127 using pulsar gating.  We use
%this position in the figures shown.

\section{Observations} 
\label{sec_obs}

We have undertaken radio observations towards PSR~J1119$-$6127 with the
Australia Telescope Compact Array (ATCA)
(\cite{fbw92}), an east-west synthesis telescope consisting
of six 22-m antennas, located
near Narrabri, NSW, Australia. These observations
are summarized in Table~\ref{chap6:tab2}; in each observation
data were taken
simultaneously at 1.4 and 2.5~GHz. The bandwidth at each
observing frequency was 128~MHz, split into 32 4-MHz channels.
Observations were carried out in pulsar-gating mode,
in which complex correlations from the antenna pairs
were recorded 32 times per pulse period, and then folded
at this period before being written to disk. 
All four Stokes parameters were recorded for each spectral channel
and pulse bin.
All observations used the same pointing center, located $\sim1'$ from
the position of the pulsar. An absolute flux density scale
was determined by observations of PKS~B1934--638, assuming
flux densities for this source of 14.9 and 11.1~Jy at
1.4 and 2.5~GHz respectively (\cite{rey94}). Both the time-dependent
gain and the polarization leakage parameters of each antenna
were determined via regular observations of PKS~B1036--697.
The final combined data set consisted of 1172860 complex
correlations spread over 60 baselines ranging between 31~m
and 5878~m.

\section{Analysis and Results} 
\label{sec_results}

\subsection{Total intensity images} 

Data were calibrated in the {\tt MIRIAD}\ package, using 
standard approaches unless otherwise stated (\cite{sk98}).
Data were first edited to eliminate portions corrupted
by interference; flux density, antenna gain and
polarization calibrations were then applied. Images of
total intensity were then formed at both 1.4 and 2.5~GHz,
using data from all pulse bins, and employing
a uniform weighting scheme to minimize side lobes. To enhance the
surface brightness sensitivity,
images were formed using only
baselines at projected spacings shorter than 7.5~k$\lambda$.
Furthermore, at 2.5~GHz we also discarded data at spacings
shorter than 0.5~k$\lambda$, to reduce the effect of
confusing side-lobes produced by the two bright nearby \HII\ regions.
The images were then deconvolved using a maximum entropy
routine (\cite{ce85}; \cite{nn86}), and
the resulting images convolved with a Gaussian restoring
beam of dimensions $29''\times25''$ (at 1.4~GHz)
and $21''\times20''$ (at 2.5~GHz), corresponding to
the diffraction-limited resolutions of the respective
data sets.

\begin{figure}
%\plotone{figures/fig2.ps}
\centerline{\psfig{file=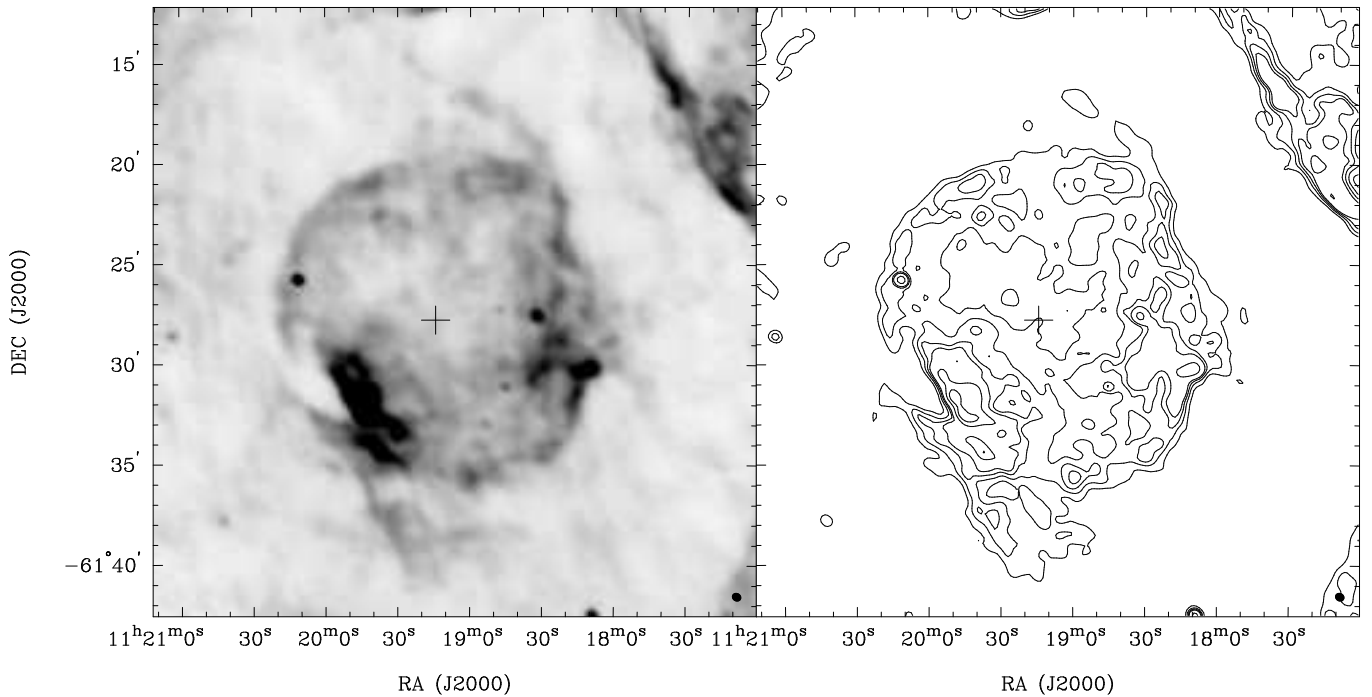,width=18cm,angle=270}}
\caption{ATCA observations of PSR~J1119--6127 at 1.4~GHz, using only
data corresponding to baselines shorter than 7.5~k$\lambda$. The greyscale
ranges from --2 to +18~mJy~beam$^{-1}$, while contours are
shown at levels of 2.5, 5, 7.5, 10, 20 and 30~mJy~beam$^{-1}$.The resolution
of the image is $29'' \times 25''$ (shown at lower right
of each panel) and the rms noise level is $\sim$ 0.5
mJy~beam$^{-1}$.}
\label{chap6:fig2}
\end{figure}

\begin{figure}
%\plotone{figures/fig3.ps}
\centerline{\psfig{file=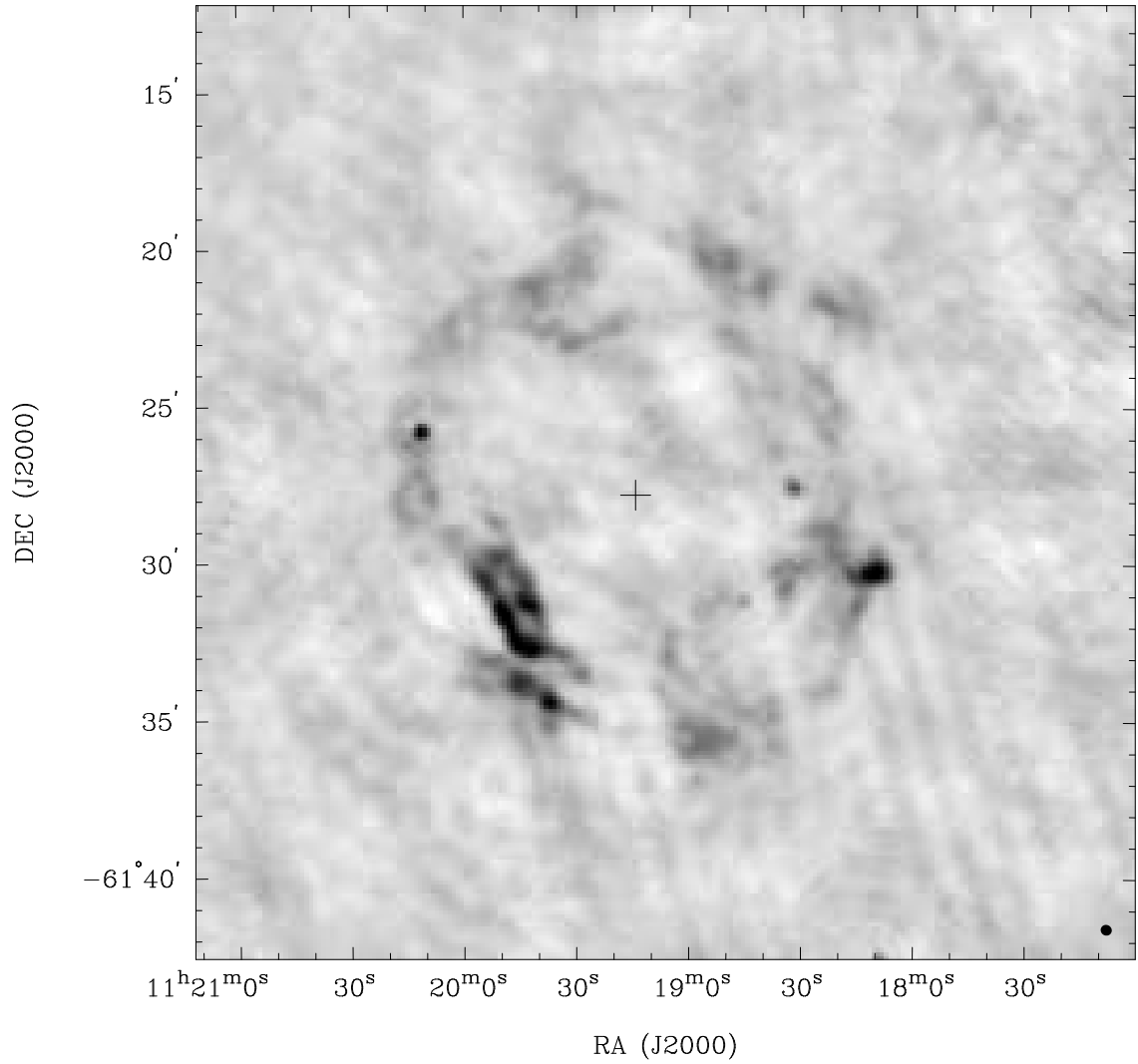,width=15cm,angle=270}}
\caption{ATCA observations of PSR~J1119--6127 at 2.5~GHz,
using data corresponding to baselines between
0.5 and 7.5~k$\lambda$. The greyscale
ranges from --2 to +10~mJy~beam$^{-1}$,
while the resolution is $21'' \times 20''$. The image quality
is significantly degraded by side-lobes from nearby bright sources.}
\label{chap6:fig3}
\end{figure}

The resulting images are shown in Figures~\ref{chap6:fig2} 
and \ref{chap6:fig3}. The rms sensitivity in each image
is $\sim$0.5~mJy~beam$^{-1}$, which is ten and five
times worse than the theoretical sensitivity expected
at 1.4~GHz and 2.5~GHz, respectively. This is not 
surprising given the very complicated nature of the region and the
presence of bright \HII\ regions in the vicinity.

In the 1.4-GHz image in Figure~\ref{chap6:fig2}, it is clear that the
extended emission seen in the lower-resolution MGPS data
shown in Figure~\ref{chap6:fig1} is now resolved into a distinct
limb-brightened elliptical shell of dimensions $14'\times16'$. 
After applying a correction for a non-zero background,
we estimate a 1.4-GHz flux density for this shell of 
$5.6\pm0.3$~Jy.
Based on its Galactic
coordinates, we designate this shell G292.2--0.5.

This shell can also be seen at 2.5~GHz in Figure~\ref{chap6:fig3};
however the image quality at this frequency is poorer than at 1.4~GHz,
because the confusing \HII\ regions are further into the wings
of the primary beam than at 1.4~GHz, and their side lobes are
consequently more difficult to deconvolve.
At 2.5~GHz the measured flux density of G292.2--0.5
is $1.6\pm0.1$~Jy. This value is a significant 
underestimate on the source's true flux density, because
at this higher frequency the largest scale to which
the interferometer is sensitive is $\sim7'$, 
implying that much of the emission from this source is not
detected in our observations.

\begin{figure}
%\plotone{figures/fig4.ps}
\centerline{\psfig{file=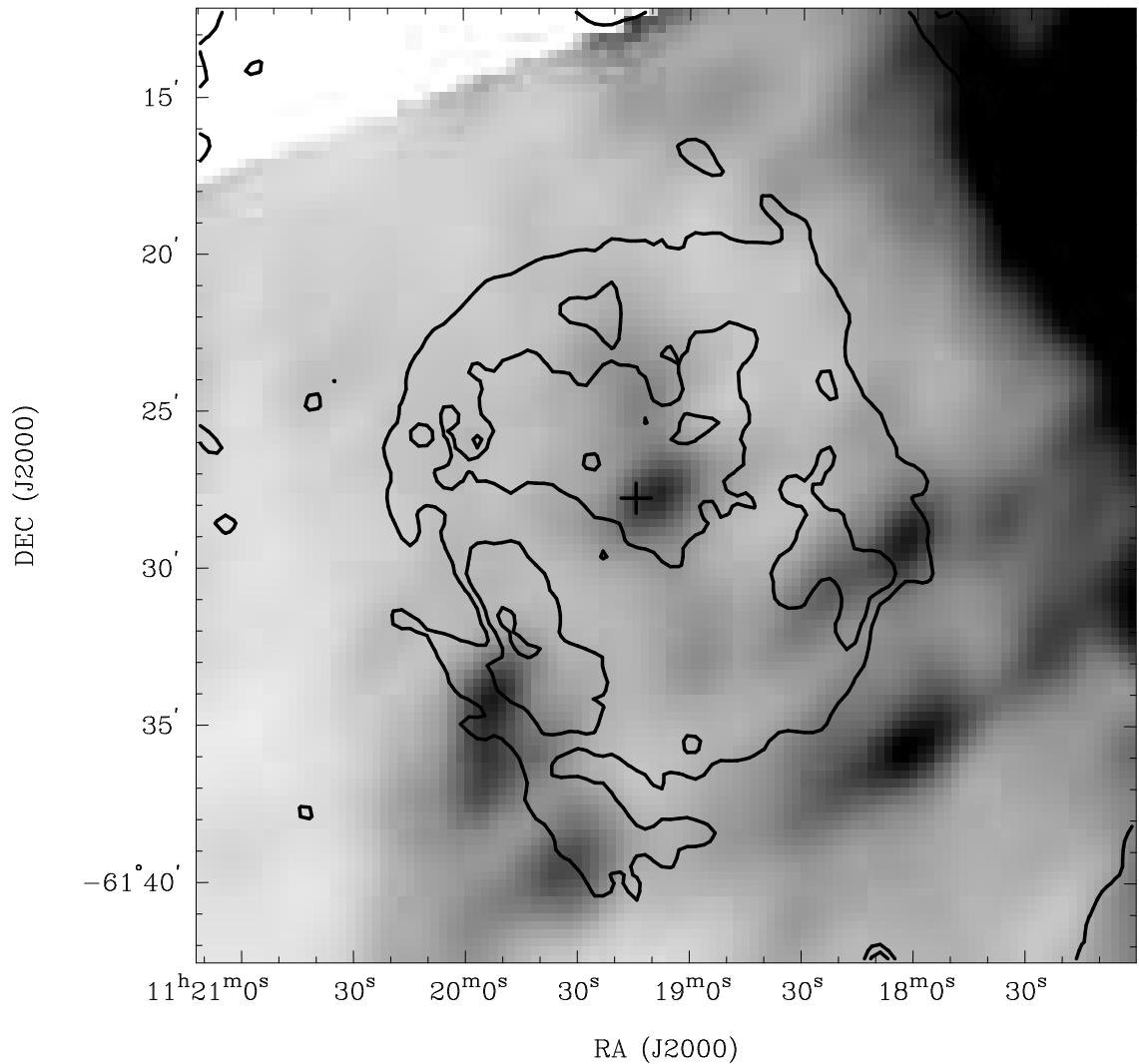,width=15cm,angle=270}}
\caption{Comparison of infrared and radio emission from
G292.2--0.5. The greyscale shows the
{\em IRAS}\ 60-$\mu$m emission from the region, taken
from the {\em IRAS}\ Galaxy Atlas (\cite{ctpb97}),
and ranging between 70 and 400~MJy~sr$^{-1}$.
Contours show 1.4 GHz radio emission as in Figure~\ref{chap6:fig2}, at levels
of 3, 10 and 30~mJy~beam$^{-1}$. The infrared source 
$\sim1'$ to the west of PSR~J1119--6127 is
IRAS~J11169$-$6111.}
\label{chap6:fig4}
\end{figure}

In Figure~\ref{chap6:fig4} we show 
an {\em IRAS}\ 60-$\mu$m image 
of the region (\cite{ctpb97}). While there is significant infrared
emission associated with the bright \HII\ regions to the
west of the pulsar, there is no detectable
emission coincident with G292.2--0.5.
The weak infrared source IRAS~J11169--6111 can be seen
close to the position of PSR~J1119--6127; this source
is probably not associated with the pulsar
(see discussion by \cite{pkc+00}).

We determined an accurate position for the pulsar as follows.
First, a delay was applied
to each frequency channel corresponding to a dispersion measure of
713~pc~cm$^{-3}$.\footnote{This was the best estimate available at the
time, and differs slightly from the published dispersion measure of
$707\pm2$~pc~cm$^{-3}$ (\cite{ckl+00}); this small difference does not
affect the results.} The data corresponding to off-pulse bins were then
subtracted from on-pulse data in the $u-v$ plane, and the position of
the pulsar determined by fitting the Fourier transform of a point
source to the resulting data set.  The resulting position and
associated uncertainty for PSR~J1119--6127 are RA~(J2000)~$11^{\rm
h}19^{\rm m}14\fs30\pm0\fs02$,
Dec.~(J2000)~$-61^{\circ}27'49\farcs5\pm0\farcs2$.  This position, marked
in all the images with a cross, puts the pulsar coincident
with the geometric center of the SNR, within the uncertainties
of such a determination.

\subsection{Spectral index determination}

A measurement of the spectral index, $\alpha$, for G292.2--0.5
(where $S_\nu \propto \nu^\alpha$) is
essential for determining whether this source is a SNR or not.
However, a spectral index cannot be determined directly from the 1.4 and
2.5~GHz images shown in Figures~\ref{chap6:fig2} and \ref{chap6:fig3}, 
because the $u-v$ coverage of the two
images differs considerably. We thus spatially filtered
the two images as follows (see also \cite{gbm+98}). 
We first corrected the 1.4~GHz
image for primary beam attenuation, and then applied the
attenuation corresponding to the shape of the
primary beam at 2.5~GHz. We then Fourier transformed
the resulting image, and re-sampled it using the transfer
function of the 2.5~GHz observations. The resulting data set
then had properties identical to the 2.5~GHz data, except that
the intensity distribution on the sky was that at 1.4~GHz.
This data set was then imaged and deconvolved in the same
way as for the 2.5 GHz data.
Both this resulting image and the
original 2.5~GHz image in Figure~\ref{chap6:fig3} 
were then smoothed to a final resolution of $30''$.

These two images could then be directly compared to determine
the spectral index of the emission. To make this determination,
we applied the approach of ``spectral tomography'' (\cite{kr97})
to G292.2--0.5.
This involved scaling the 2.5~GHz image by a trial spectral index
$\alpha_t$, and then 
subtracting this scaled image from the 1.4~GHz image, to form a difference
image $I_{\alpha_t}$, defined by:

\begin{equation} 
I_{\alpha_t} = I_{1.4} - \left( \frac{1.4}{2.5} \right)^{\alpha_t} I_{2.5},
\end{equation} 
where $I_{1.4}$ and $I_{2.5}$ are the images to be compared at
1.4 and 2.5~GHz respectively. The spectral index, $\alpha$, of
a particular feature is simply the
value of $\alpha_t$ at which its emission blends into the background.
An uncertainty in this spectral index is estimated by finding
the range in values of $\alpha_t$ at which the residual
at this position in the difference image becomes significant.

\begin{figure}
\centerline{\psfig{file=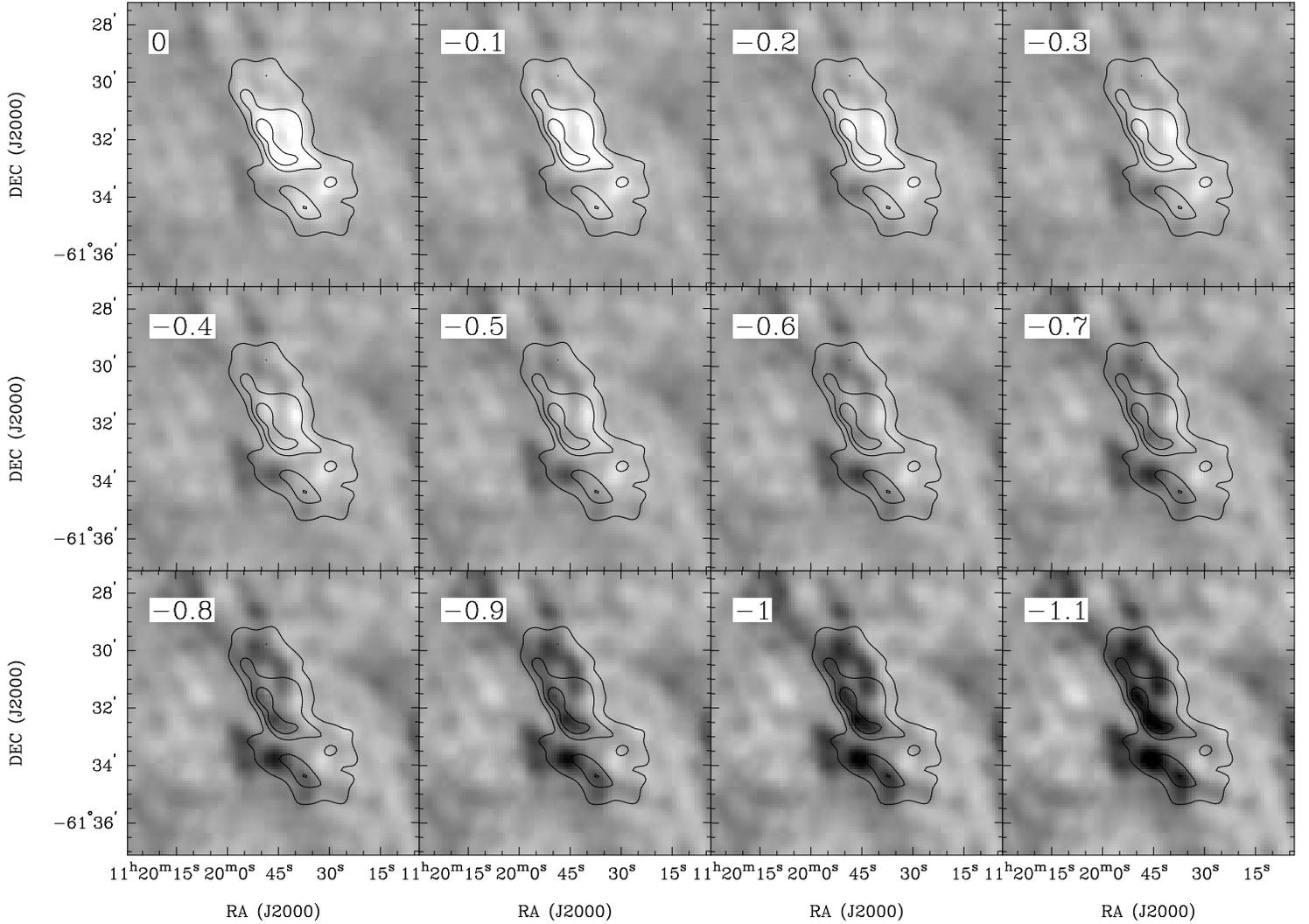,height=14cm,angle=270}}
\caption{Tomographic spectral index images for the bright
south-eastern rim of the shell G292.2--0.5. The greyscale
shows a series of difference images between 1.4 and 2.5 GHz
data which have been matched in $u-v$ coverage.
For each panel, the trial spectral index is shown at upper left.
The contours in each panel correspond to the
1.4 GHz data shown in Figure~\ref{chap6:fig2}, 
at levels of 10, 20 and 30~mJy~beam$^{-1}$.
Light (dark) regions of the image indicate
that a feature has a more negative (positive) spectral index
than the trial value. The transition from light to dark
occurs at a spectral index $\sim -0.6$.}
\label{chap6:fig5}
\end{figure}

In Figure~\ref{chap6:fig5} we show a series of such difference 
images for the region surrounding the bright south-eastern rim
of G292.2--0.5, for trial spectral indices
in the range $0.0 > \alpha_t > -1.1$ at intervals $\Delta\alpha_t = 0.1$.
For a trial 
spectral index $\alpha_t = 0$, the significant negative residuals
indicate that the emission from the shell has a non-thermal
($\alpha < 0$) spectrum. By finding the trial spectral index
at which the residuals make the transition
from negative to positive, we estimate that the
spectral index of this part of
G292.2--0.5 is $\alpha = -0.6 \pm 0.2$. Other parts of the
shell are significantly fainter at 2.5~GHz, and a spectral index
determination is correspondingly more difficult. However,
results for these regions also suggest a spectral
index $\alpha \sim -0.6$.

\subsection{Polarization}

SNRs are significantly linearly polarized, typically at
a fractional level of 10--20\%. Thus
the presence or absence of linear polarization 
is a further useful discriminant as to whether a source is a SNR.
Unfortunately there is
a significant off-axis response in linear polarization,
meaning that instrumental artifacts due to the bright \HII\ regions
NGC~3603 and NGC~3576
dominate emission in Stokes~$Q$ and $U$ at the
position of G292.2--0.5. We were thus unable to make
any measurement of the polarization properties of this object.

\subsection{Limits on emission from a radio PWN}

There is no obvious emission at the pulsar's position
in either of Figures~\ref{chap6:fig2} or \ref{chap6:fig3} which
might be associated with a PWN. In order to
put quantitative limits on the surface brightness of
such a source, we carried out the following steps.

First, we excluded pulse bins corresponding to times when the pulsar
was ``on'', emission from which could potentially mask emission from a
coincident PWN (\cite{gsfj98}; \cite{sgj99}). Next, we subtracted 
a model of the emission from the shell itself and from
external sources from the $u-v$ data.
The resulting image had greatly improved sensitivity
to emission at or near the pulsar's position (see \cite{gbs00}).  We
then carried out a series of trials, stepping through all angular sizes
between the resolution of the data and the extent of G292.2--0.5.  For
each trial, we simulated a PWN by adding to the $u-v$ data the Fourier
transform of a faint circular disk centered on the pulsar, and with
diameter equal to the trial angular size. We then made an image of the
region, convolving it with a Gaussian of FWHM slightly smaller than the
size of the disk to give maximal sensitivity to structures on that
scale. We repeated this process at increasing flux
densities for the simulated PWN, until it could be detected in the
resulting image at the $5\sigma$ level (e.g.\ \cite{gs92};
\cite{gbs00}).  The surface brightness of this marginally detectable
disk was taken as our sensitivity to a PWN at a given angular size.

Figure~\ref{chap6:fig6} shows a plot of the corresponding 1.4 and 2.5~GHz
upper limits on surface brightness of any PWN, as a function of
angular size. In both cases, we have scaled our results to
an observing frequency of 1~GHz by assuming a PWN
spectral index $\alpha = -0.3$ as seen for the Crab Nebula.

\section{An association between G292.2--0.5 and PSR~J1119--6127?}
\label{sec_assoc}

In order for a source to be classified as a shell-type
SNR, it needs to have a limb-brightened morphology, show
a non-thermal spectral
index ($\alpha \sim -0.5$), have a high
radio to infrared flux ratio, and be linearly
polarized. We have shown that G292.2--0.5 meets
all of the first three criteria (as explained in Section~\ref{sec_results},
polarization measurements were not possible for this source
due to instrumental effects). We therefore conclude
that G292.2--0.5 is a previously unidentified SNR.

The 1-GHz surface brightness of this SNR\footnote{using the usual
definition $\Sigma_\nu \equiv S_\nu/\theta^2$, where
$\theta$ is the angular diameter of the SNR, and where
the effects of limb-brightening have been disregarded}
is $\Sigma \sim4.7\times10^{-21}$~W~m$^{-2}$~Hz$^{-1}$~sr$^{-1}$. 
This value is not particularly faint, being of order the approximate
completeness limit of the entire sample of Galactic SNRs (\cite{gj95c}). Thus
the fact that this SNR had until now gone undetected is not
a result of lack of sensitivity, but is entirely due to the severe
selection effects associated with identifying SNRs in complicated
regions. Undoubtedly there remain many other reasonably bright
SNRs which are similarly hidden beneath the side-lobes from
adjacent strong sources.

Associations between pulsars and SNRs are usually judged on agreement
in distance, agreement in age, the space velocity inferred
from the offset of the pulsar with respect to the SNR's apparent center,
and the likelihood of a chance coincidence between the two
sources. We now consider each of these criteria in turn.

As discussed by Camilo \etal (2000a\nocite{ckl+00}), the distance
inferred for PSR~J1119--6127 from its dispersion measure is $>$30~kpc,
and is certainly a severe overestimate. Meanwhile, no distance
estimate is possible for SNR~G292.2--0.5 from the available data.
Thus we are unable to determine whether distances to
the two sources are comparable.
Camilo \etal (2000a\nocite{ckl+00}) note that the Carina spiral
arm of the Galaxy crosses the line of sight at distances
of 2.4 and 8~kpc, and therefore assume that the pulsar's
true distance falls somewhere in this range. We make a similar
assumption for SNR~G292.2--0.5, and in future discussion
denote the distance to the source as $5d_5$~kpc. The
radius of the SNR is then $R=(10.9\pm0.7)d_5$~pc.

We can derive an approximate age for SNR~G292.2--0.5 as follows.
For a uniform ambient medium of pure hydrogen with density $n_0$~cm$^{-3}$,
the mass swept up by the SNR is $\sim140n_0d_5^3$~$M_{\sun}$.
For typical ejected masses and ambient densities, the SNR is
then partly in transition to the adiabatic
(Sedov-Taylor) phase of expansion
(see e.g.\ \cite{dj96}). We can therefore
use the expected rate of expansion 
in the adiabatic phase to derive an upper limit on the SNR age 
of $(7\pm1) \left(n_0/E_{51} \right)^{1/2}d_5^{5/2}$~kyr,
where $E_{51}$ is the kinetic energy of the explosion in units of
$10^{51}$~erg.  For a typical value $n_0/E_{51}=0.2$ (\cite{fgw94}), we find
that the age of the SNR must be less than $\sim3d_5^{5/2}$~kyr. This is in
good agreement with the upper limit on the pulsar's age of
$1.7\pm0.1$~kyr estimated by Camilo \etal (2000a\nocite{ckl+00})
from pulsar timing. 

We estimate that the pulsar is offset by no more than $\sim1'$
from the geometric center of the SNR. If this geometric
center corresponds to the site of the supernova explosion,
and if the pulsar and SNR are physically associated, then
a typical transverse velocity for the pulsar of 
$500V_{500}$~\kms\ implies an upper limit on
the age for the system of $<2.9d_5/V_{500}$~kyr,
consistent with the ages estimated
separately for the pulsar and the SNR.

As discussed above, G292.2--0.5 has a surface brightness comparable to
the completeness limit of the Galactic SNR population. Thus a good
estimate of the probability of a chance alignment between the pulsar
and the SNR can be made by considering the distribution of known SNRs down to
this surface brightness limit.  In a representative region of the
Galactic Plane covering the range $270^\circ \le l \le 330^\circ$, $|b|
\le 2^\circ$, there are 20 other SNRs brighter than G292.2--0.5
(\cite{wg96}; \cite{gre00}).
The probability that such a SNR should have its center
lying by chance within $1'$ of PSR~J1119--6127 is therefore
$\sim7\times10^{-5}$.  Even if the geometric center of a SNR does not
correspond to its true center, the offset between the two is probably
no more than $\sim$25\% of the SNR's radius (e.g.\ \cite{dj96};
\cite{zan00}), so that the probability of a chance alignment is still
$\ll 10^{-3}$.

To summarize, we find that SNR~G292.2--0.5 is of comparable age to
PSR~J1119--6127, that the offset of the pulsar from the SNR's center is
consistent with the youth of the system, and that the probability is
low that the two sources align so closely by chance. We therefore
conclude that it is highly likely that PSR~J1119--6127
and SNR~G292.2--0.5 are physically associated.

\section{Implications of the non-detection of a radio PWN}
\label{sec_pwn}

\subsection{PWNe around young pulsars}

The Crab pulsar is by far the best-studied young pulsar.
In the evolutionary picture inferred for this pulsar
and its associated nebula, 
a young pulsar is born spinning rapidly
($P \la 20$~ms).  Over the next thousand years, the pulsar
undergoes only modest spin-down so that it is still spinning rapidly
and has a high spin-down luminosity. The interaction of the resulting
pulsar wind with the ambient medium produces a radio-bright PWN. The
other two rapidly spinning pulsars in Table~\ref{chap6:tab1}, 
PSRs~B0540--69 and J0537--6910, also have these properties.

However, the other three young pulsars in Table~\ref{chap6:tab1},
namely PSRs~J1846--0258, B1509--58 and J1119--6127, are all of a comparable
age to the Crab-like pulsars, but spin much more slowly ($P>100$~ms)
and have much higher magnetic fields ($B>10^{13}$~G).  If, like the
Crab-like pulsars, these high-field pulsars were born spinning rapidly,
then we can infer that they lost most of their rotational energy early
in their lives. Given this important difference in their evolutionary
histories, it is perhaps not reasonable to expect that they should
power radio-bright PWNe like those produced by the Crab-like pulsars.
While PSR~J1846--0258 is associated with a radio PWN (\cite{bh84};
\cite{gvbt00}), no radio PWN has been detected around PSR B1509--58
(\cite{gbm+98}), although the sensitivity to such a source is
limited by the complicated nature of the region. Furthermore, we have shown
here that to even more stringent limits, no PWN is present around
PSR~J1119--6127 either.

\begin{figure}
%\plotone{figures/fig6.ps}
\centerline{\psfig{file=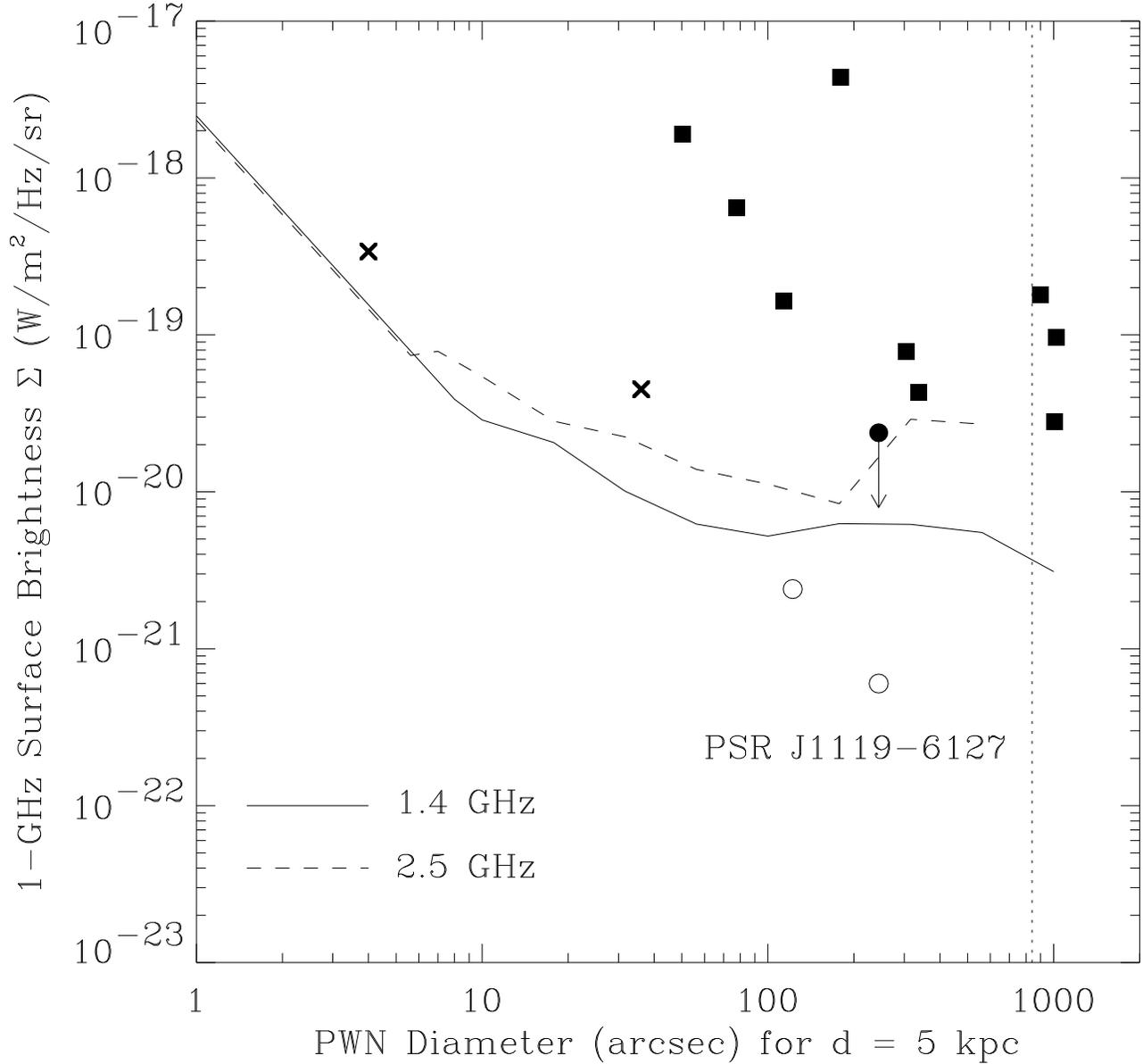,width=18cm}}
\caption{1~GHz surface brightness vs diameter for
known radio PWNe, and corresponding upper limits for 
a PWN associated with PSR~J1119--6127. 
Plotted points correspond to the properties
of PWNe as listed in Table~\ref{chap6:tab3}, scaled
to a distance of 5~kpc. Squares
indicate static nebulae, crosses indicate
nebulae with a bow-shock morphology, while the solid
circle indicates the upper limit on a radio PWN
associated with PSR~B1509--58.
The open circles indicate the predicted size and surface
brightness for a PWN associated with PSR~J1119--6127 using
cases NE (lower right) and case E (upper left) of
the PWN model of RC84 (see text for details);
the solid and dashed lines respectively correspond to the 1.4 and
2.5~GHz observed upper limits on such emission.
The vertical dotted line
indicates the diameter of G292.2--0.5; this is
a hard upper limit on the extent of any PWN associated with
PSR~J1119--6127.}
\label{chap6:fig6}
\end{figure}

In Table~\ref{chap6:tab3} we list the properties of a representative
group of radio PWN. We plot the 
resulting sizes and 1-GHz surface brightnesses of these sources
in Figure~\ref{chap6:fig6},
all scaled to a common distance of 5~kpc. Two types of
PWN are shown: those with a ``static'' morphology,
in which the PWN is confined by the gas pressure of the
surrounding medium (e.g.\ \cite{aro83}; \cite{rc84}, hereafter RC84), 
and those with a ``bow-shock''
morphology, where the PWN is confined by ram pressure
resulting from the pulsar's motion with respect to its
environment (e.g.\ \cite{fggd96}). It can be seen from the upper
limits derived in Section~\ref{sec_results} and also
plotted in Figure~\ref{chap6:fig6} that 
each of the PWNe considered
would have been easily detected in our observations.

In the case of PSR~B1509--58, we plot the upper limit of
Gaensler \etal (1999\nocite{gbm+98}). The angular
size for this PWN has been
estimated by assuming that the PWN has freely
expanded since the pulsar's birth with an expansion
velocity $V = 1700$~\kms\ as
measured for the Crab Nebula (\cite{tri68}). It
can be seen that the upper limits on the brightness
of a PWN around PSR~J1119--6127 are ten times
more stringent than those for PSR~B1509--58, and thus
are potentially a better test of the evolution of
high-magnetic field pulsars.

\subsection{A simple model for PWN evolution}

An explanation for the absence of a radio PWN around PSR~B1509--58 
was proposed by Bhattacharya~(1990)\nocite{bha90}, who pointed
out that even if a high magnetic-field pulsar is born spinning rapidly, it
will quickly slow down to a long period due to severe magnetic
braking. At later times, there is therefore no significant
energy injection into the PWN from the pulsar,
and the energetics of the nebula are dominated
by the significant losses which it 
experiences due to expansion into the ambient medium. The net
result is that a PWN associated with a high-field pulsar
should have an observable lifetime which is very brief,
corresponding to the reduced period for which the pulsar
provides significant amounts of energy to the nebula.

%The nebula undergoes a rapid decay in luminosity and becomes
%undetectable early in its lifetime. The net energy loss rate of the
%nebula $\dot{E}_{\rm PWN}$ can be modeled as the difference between
%the energy injection rate from the pulsar $\dot{E}_{\rm psr}$ and the
%energy loss rate from expansion $\dot{E}_{\rm expansion}$.
%
%The rate of energy loss from the pulsar as a function of time $t$,
%assuming magnetic dipole radiation, is (\cite{ps73})
%
%\begin{equation} 
%\dot{E}_{\rm psr} = \frac{L_{0}}{\left( 1 + t/\tau_{0} \right)^{2}} \, ,
%\end{equation} 
%
%\noindent
%where $L_{0} \equiv 4 \pi^{2} I \dot{P}_{0} / P_{0}^{3}$ is the
%initial energy loss rate of the pulsar and $\tau_{0} \equiv P_{0} / 2
%\dot{P}_{0}$. Here $P_{0}$ and $\dot{P}_{0}$ are the period and period
%derivative of the pulsar at birth. The energy loss rate $\dot{E}_{\rm
%expansion}$ from the expansion of the nebula as a function of time $t$
%(assuming uniform expansion in time) is
%
%\begin{equation}
%\dot{E}_{\rm expansion} = -\frac{E_{\rm PWN}}{t} \, ,
%\end{equation} 
%
%\noindent
%where $E_{\rm PWN}$ is the energy contained in the PWN. These
%equations can be combined to obtain an expression for the evolution of
%the energy of the PWN as a function of time:
%
%\begin{equation} 
%\dot{E}_{\rm PWN} = 
%-\frac{E_{\rm PWN}}{t} 
%+\frac{L_{0}}{\left( 1 + t/\tau_{0} \right)^{2}} .
%\end{equation} 
%
%\noindent

The radio luminosity evolution of a PWN has been considered
in detail by several authors (e.g.\ \cite{ps73}; RC84).
In these studies, it is found that the evolution of a radio PWN
can be divided into two main stages. The transition
occurs at a time $t \approx \tau_0 \equiv P_0/(n-1)\dot{P}_0$, where
$P_0$ and $\dot{P}_0$ are the initial period and period derivative
of the pulsar, respectively, and $n$ is the pulsar's braking
index (assumed to not vary with time).

In the first stage of evolution ($t \la \tau_0$),
the pulsar is a significant source of energy
for the expanding PWN. For times
immediately preceding $t = \tau_0$,
the radio luminosity of the PWN at a given frequency 
decreases with time as $L_\nu(t) \propto t^{(7\alpha-3)/4}$ (RC84), 
where $\alpha$ is the radio spectral index of the PWN
defined by $S_\nu \propto \nu^{\alpha}$.
The radio luminosity decreases, 
despite the fact that the pulsar continues to inject energy into the
PWN, because of losses resulting from expansion of the nebula.

In the succeeding stage of evolution ($t \ga \tau_{0}$), the pulsar 
has now transferred the bulk of its initial rotational
energy into the PWN, and no longer injects significant
amounts of energy into it. The radio luminosity
of the PWN now decreases more steeply than in the first
stage, so that $L_\nu \propto t^{-2\gamma}$, where
$\gamma = 1-2\alpha$ is the energy-index of the
injected particle spectrum.

RC84 consider two
cases for the evolution of a PWN: case ``NE'', in
which the energies in relativistic electrons and magnetic
fields evolve independently, and case ``E'', in
which there is equipartition between particles and magnetic
fields. 

Analytic solutions for the evolution
of the PWN are only possible for case NE. In this case,
we find that at a frequency $\nu$, and at a time
$t > \tau_{0}$,\footnote{which is the only time-range
of interest for the pulsars being considered here} 
the spectral luminosity of the nebula is given by (RCW84):
\begin{equation} 
L_\nu = K\, L_0^{(7-\gamma)/8}\, \tau_0^{(1+9\gamma)/8}\, (p-\gamma)^{-1}\, t^{-2\gamma}
\end{equation} 
where $L_0 \equiv 4\pi^2I \dot{P}_0/P_0^3$ is the
initial spin-down luminosity of the pulsar, $p = (n+1)/(n-1)$ and
$K$ is a constant of proportionality.
$K$ depends on the mass of the slowly-moving ejecta
and on the details of the injected particle spectrum (RC84) ---
we assume that these quantities are the same for each of
the pulsars we are considering.

The ratio of spectral luminosities
$L_{\nu,1}$ and $L_{\nu,2}$ for two PWNe with ages $t_{1}$ and $t_{2}$
respectively, and with similar spectral indices, is then:
\begin{equation} 
\frac{L_{\nu,1}}{L_{\nu,2}} = \left( 
\frac{L_{1}}{L_{2}} \right)^{(7-\gamma)/8}
\left( \frac{ \tau_{1}}{\tau_{2}} \right)^{(1+9\gamma)/8}
\left( \frac{ p_1-\gamma}{p_2-\gamma} \right)^{-1}
\left( \frac{t_{1}}{t_{2}} \right)^{-2\gamma} ,
\end{equation} 
where $L_{1,2}$, $\tau_{1,2}$ and $p_{1,2}$
are the initial spin-down luminosities $L_0$, the
``initial characteristic ages'' $\tau_{0}$ 
and the parameter $p=(n+1)/(n-1)$ for the two pulsars, respectively.

Since the initial surface magnetic field has the
dependency $B_0 \propto (P_0 \dot{P}_0)^{1/2}$, we
can write $\tau_0 \equiv P_0 / (n-1) \dot{P}_0 \propto P_0^2/(n-1)B_0^2$.
Similarly, we find that $L_0 \propto \dot{P}_0 /P_0^3 \propto P_0^2/B_0^4$, so that:

\begin{equation} 
\frac{L_{\nu,1}}{L_{\nu,2}} = 
\left( \frac{P_{1}}{P_{2}} \right)^{\frac{-13+11\gamma}{4}}
\left( \frac{B_{1}}{B_{2}} \right)^{\frac{3-5\gamma}{2}}
\left( \frac{n_1-1}{n_2-1} \right)^{-\frac{1+9\gamma}{8}}
\left( \frac{ p_1-\gamma}{p_2-\gamma} \right)^{-1}
\left( \frac{t_{1}}{t_{2}} \right)^{-2\gamma} ,
\end{equation} 
where $P_{1,2}$, $B_{1,2}$ and $n_{1,2}$ are the initial periods, initial
magnetic fields and braking indices of the two pulsars. 

We now convert spectral luminosities into 1-GHz
surface brightnesses using the fact that $\Sigma \propto L/(Vt)^2$, where
we have assumed that the PWN is freely expanding. Assuming
that $\alpha = -0.3$ as for the
Crab Nebula, we have that $\gamma = 1-2\alpha = 1.6$,
so that:
\begin{eqnarray}
\frac{\Sigma_{1}}{\Sigma_{2}} =
\left(  \frac{P_{1}}{P_{2}} \right)^{+1.15} 
\left(  \frac{B_{1}}{B_{2}} \right)^{-2.50} 
\left(  \frac{n_1-1}{n_2-1} \right)^{-1.93}
\left( \frac{ p_1-\gamma}{p_2-\gamma} \right)^{-1}  \nonumber \\
%\left( \frac{ p_1-\gamma}{p_2-\gamma} \right)^{-1}  
\left(  \frac{V_{1}}{V_{2}} \right)^{-2.0} 
\left(  \frac{t_{1}}{t_{2}} \right)^{-5.2} . 
\label{chap6:surfbright}
\end{eqnarray}

By making the assumptions that $P_0$ and $V$ for PSR~J1119--6127 and
its associated PWN are similar to the corresponding values for the Crab, that
the characteristic age of PSR~J1119--6127
is a good approximation to its true age, and
that the pulsar's surface magnetic field does not significantly
evolve with time, we can then use
Equation~(\ref{chap6:surfbright}) to predict the brightness of a radio
PWN associated with PSR~J1119--6127 for case NE.  Using the parameters listed in
Table \ref{chap6:tab4}, we predict a 1~GHz surface brightness for a
radio PWN of $\Sigma \simeq 6 \times
10^{-22}$~W~m$^{-2}$~Hz$^{-1}$~sr$^{-1}$.  It can be seen from
Figure~\ref{chap6:fig6} that this predicted value is significantly fainter
than the surface brightness of any detected radio PWN, and is
consistent with the upper limits derived in Section~\ref{sec_results}.

No analytic solution is possible for case E. However, Figure~2
of RC84 shows that the main difference
between cases NE and E is that the radius of the expanding
PWN is $\sim2$ times smaller in the latter case.
The predicted surface brightness (plotted in Figure~\ref{chap6:fig6}) is
then $\sim4$ times larger than for case NE, but it still below the 
observed upper limits. 

Note that while the direct dependency of radio luminosity on initial
period in Equation~(\ref{chap6:surfbright}) is weak, a longer initial
period would also imply that the pulsar's true age was smaller than its
characteristic age --- the strong dependence of $\Sigma$ on $t$ must
then also be taken into account.  For example, Camilo \etal
(2000a\nocite{ckl+00}) show that for an initial period $P_0 = 200$~ms,
the true age for PSR~J1119--6127 would only be 1.2~kyr.  In this case,
the parameters in Table~\ref{chap6:tab4} would then cause us to
underestimate the surface brightness by a factor $\sim100$, and
would result in a predicted surface brightness well above
our sensitivity limits. Our model thus argues in favor of a small
initial period  for PSR~J1119--6127.

We can also apply this model to PSR~B1509--58, for
which we predict a 1-GHz
surface brightness $\Sigma \sim 8
\times 10^{-21}$~W~m$^{-2}$~Hz$^{-1}$~sr$^{-1}$,
in agreement with the upper limit set
by Gaensler \etal (1999\nocite{gbm+98}). 
For the apparently younger PSR~J1846--0258, Equation~(\ref{chap6:surfbright})
predicts $\Sigma \sim7\times10^{-20}$~W~m$^{-2}$~Hz$^{-1}$~sr$^{-1}$.
While this is  approximately a factor of 2 fainter than the observed
surface brightness of the associated PWN, this source
has a significantly different spectral index
than the Crab Nebula ($\alpha = 0$; \cite{bh84}), and so it
is not reasonable to apply a simple scaling from the properties
of the Crab as has been done here.
Furthermore, while braking indices have been
measured for PSRs~B1509--58
and J1119--6127 (\cite{kms+94}; \cite{ckl+00}), no such measurement
has yet been made for PSR~J1846--0258,
and its characteristic age may thus not be a good approximation to
its true age.

%We thus find that using a simple model for the evolution
%of a radio PWN, the absence of a detectable radio PWN
%around PSR~B1119--6127 can be understood in terms
%of the short lifetime of such objects when powered by
%high magnetic-field pulsars, as originally
%proposed by Bhattacharya (1990\nocite{bha90}).

\section{Conclusions} 

Radio observations of the young pulsar J1119--6127
have revealed a limb-brightened radio shell, G292.2--0.5,
of diameter $\sim15'$.
The spectral index $\alpha = -0.6\pm 0.2$ measured for
this shell, together with its lack of infrared emission,
argue that G292.2--0.5 is a previously unidentified SNR.
The small age inferred for G292.2--0.5, together with the proximity of its
geometric center to the position of PSR~J1119--6127, argue
for a physical association between the two objects.

Radio emission from a pulsar-powered nebula was not detected
in our observations. We have shown that our observed
upper limits are consistent with the expectation that
the pulsar spun down rapidly to long periods and dumped the bulk of
its energy into the nebula at early times. The energy losses from
subsequent expansion of the nebula have caused the
radio emission to rapidly fade so that it
is now well below the detection threshold of our observations.

The presence of a pulsar within a SNR is usually inferred
either directly, via detection of pulsations, or indirectly,
through identification of a pulsar wind nebula within the SNR. 
The fact that both PSRs~B1509--58 and J1119--6127 are
faint radio pulsars with no detectable radio PWNe
suggests that many other young SNRs 
which have only been studied at radio wavelengths
could similarly  contain pulsars which are too faint to 
detect (or which are beaming away from us), and which do
not power detectable radio PWNe. 
The failure to detect radio pulsars and/or PWNe in many SNRs could
thus possibly be explained if pulsars with high magnetic
fields comprise a significant fraction of the pulsar
population.
In X-rays, young pulsars generally have higher luminosities,
broader beams and more prominent nebulae than in the radio,
so that imaging these
systems at higher energies
might be a better way to infer their presence.

%Although this explanation may well account for the absence of PWNe in
%young shell SNRs, whether it can account for all such absences can
%only be answered by detailed population synthesis modeling to
%determine whether the properties of the entire pulsar population are
%consistent with the existence of a much larger population of young
%high-magnetic-field pulsars than has been previously supposed.

\begin{acknowledgements}

We thank the referee, Steve Reynolds, for his many useful
corrections and suggestions, and for his patient explanations of the
results from RC84.  
The Australia Telescope is funded by the Commonwealth of Australia for
operation as a National Facility managed by CSIRO.  
B.M.G. acknowledges the support of NASA through Hubble
Fellowship grant HST-HF-01107.01-A awarded by the Space Telescope
Science Institute, which is operated by the Association of Universities
for Research in Astronomy, Inc., for NASA under contract NAS 5--26555.
F. Crawford and V.M.K. were partly supported by a National Science Foundation
Career Award (AST-9875897).
F. Camilo acknowledges the support of NASA grant NAG~5-3229.
This research has made use of NASA's Astrophysics Data System Abstract
Service and of the SIMBAD database, operated at CDS, Strasbourg,
France.

\end{acknowledgements}

\bibliographystyle{apj1}
\bibliography{journals,modrefs,psrrefs,crossrefs}

\begin{thebibliography}{}

\bibitem[Arons 1983]{aro83}
Arons, J. 1983, { Nature}, {\rm 302}, 301.

\bibitem[Becker \& Helfand 1984]{bh84}
Becker, R.~H. \& Helfand, D.~J. 1984, { ApJ}, {\rm 283}, 154.

\bibitem[Bhattacharya 1990]{bha90}
Bhattacharya, D. 1990, { J. Astrophys. Astr.}, {\rm 11}, 125.

\bibitem[Brazier \& Johnston 1999]{bj98}
Brazier, K. T.~S. \& Johnston, S. 1999, { MNRAS}, {\rm 305}, 671.

\bibitem[Camilo \etal  2000a]{ckl+00}
Camilo, F., Kaspi, V.~M., Lyne, A.~G., Manchester, R.~N., Bell, J.~F., D'Amico,
  N., McKay, N. P.~F., \& Crawford, F. 2000a, { ApJ}, {\rm 541}, 367.

\bibitem[Camilo \etal  2000b]{clm+00}
Camilo, F. \etal  2000b, in { Pulsar Astronomy --- 2000 and Beyond, {IAU}
  Colloquium 177}, ed.\ M. Kramer, N. Wex, \& R. Wielebinski, (San Francisco:
  Astronomical Society of the Pacific), 3.

\bibitem[Cao \etal  1997]{ctpb97}
Cao, Y., Terebey, S., Prince, T.~A., \& Beichman, C.~A. 1997, { ApJS}, {\rm
  111}, 387.

\bibitem[Cornwell \& Evans 1985]{ce85}
Cornwell, T.~J. \& Evans, K.~F. 1985, { A\&A}, {\rm 143}, 77.

\bibitem[De~Pree, Nysewander, \& Goss 1999]{dng99}
De~Pree, C.~G., Nysewander, M.~C., \& Goss, W.~M. 1999, { AJ}, {\rm 117}, 2902.

\bibitem[Dohm-Palmer \& Jones 1996]{dj96}
Dohm-Palmer, R.~C. \& Jones, T.~W. 1996, { ApJ}, {\rm 471}, 279.

\bibitem[{Frail} \etal  1996]{fggd96}
{Frail}, D.~A., {Giacani}, E.~B., {Goss}, W.~M., \& {Dubner}, G. 1996, { ApJ},
  {\rm 464}, L165.

\bibitem[Frail, Goss, \& Whiteoak 1994]{fgw94}
Frail, D.~A., Goss, W.~M., \& Whiteoak, J. B.~Z. 1994, { ApJ}, {\rm 437}, 781.

\bibitem[Frater, Brooks, \& Whiteoak 1992]{fbw92}
Frater, R.~H., Brooks, J.~W., \& Whiteoak, J.~B. 1992, { J. Electr. Electron.
  Eng. Aust.}, {\rm 12}, 103.

\bibitem[Gaensler, Bock, \& Stappers 2000]{gbs00}
Gaensler, B.~M., Bock, D. C.-J., \& Stappers, B.~W. 2000, { ApJ}, {\rm 537},
  L35.

\bibitem[Gaensler \etal  1999]{gbm+98}
Gaensler, B.~M., Brazier, K. T.~S., Manchester, R.~N., Johnston, S., \& Green,
  A.~J. 1999, { MNRAS}, {\rm 305}, 724.

\bibitem[Gaensler \& Johnston 1995]{gj95c}
Gaensler, B.~M. \& Johnston, S. 1995, { MNRAS}, {\rm 277}, 1243.

\bibitem[Gaensler \etal  1998]{gsfj98}
Gaensler, B.~M., Stappers, B.~W., Frail, D.~A., \& Johnston, S. 1998, { ApJ},
  {\rm 499}, L69.

\bibitem[Gotthelf \& Vasisht 2000]{gv00}
Gotthelf, E.~V. \& Vasisht, G. 2000, in { Pulsar Astronomy --- 2000 and Beyond,
  {IAU} Colloquium 177}, ed.\ M. Kramer, N. Wex, \& R. Wielebinski, (San
  Francisco: Astronomical Society of the Pacific), 699.

\bibitem[Gotthelf \etal  2000]{gvbt00}
Gotthelf, E.~V., Vasisht, G., Boylan-Kolchin, M., \& Torii, K. 2000, { ApJ},
  {\rm 542}, L37.

\bibitem[Green \etal  1999]{gcl98}
Green, A.~J., Cram, L.~E., Large, M.~I., \& Ye, T. 1999, { ApJS}, {\rm 122},
  207.
\newblock (http://www.astrop.physics.usyd.edu.au/MGPS/).

\bibitem[Green 2000]{gre00}
Green, D.~A. 2000, { A {C}atalogue of {G}alactic {S}upernova {R}emnants (2000
  {A}ugust {V}ersion)}, (Cambridge: Mullard Radio Astronomy Observatory).
\newblock (http://www.mrao.cam.ac.uk/surveys/snrs/).

\bibitem[Green \& Scheuer 1992]{gs92}
Green, D.~A. \& Scheuer, P. A.~G. 1992, { MNRAS}, {\rm 258}, 833.

\bibitem[Kaspi 1998]{kas98}
Kaspi, V.~M. 1998, in { Neutron Stars and Pulsars: Thirty Years after the
  Discovery}, ed.\ N. Shibazaki, N. Kawai, S. Shibata, \& T. Kifune, (Tokyo:
  Universal Academy Press), p.~401.

\bibitem[Kaspi \etal  1994]{kms+94}
Kaspi, V.~M., Manchester, R.~N., Siegman, B., Johnston, S., \& Lyne, A.~G.
  1994, { ApJ}, {\rm 422}, L83.

\bibitem[Katz-Stone \& Rudnick 1997]{kr97}
Katz-Stone, D.~M. \& Rudnick, L. 1997, { ApJ}, {\rm 488}, 146.

\bibitem[Lyne \etal  2000]{lcm+00}
Lyne, A.~G. \etal  2000, { MNRAS}, {\rm 312}, 698.

\bibitem[Marshall \etal  1998]{mgz+98}
Marshall, F.~E., Gotthelf, E.~V., Zhang, W., Middleditch, J., \& Wang, Q.~D.
  1998, { ApJ}, {\rm 499}, L179.

\bibitem[Narayan \& Nityananda 1986]{nn86}
Narayan, R. \& Nityananda, R. 1986, { Ann. Rev. Astr. Ap.}, {\rm 24}, 127.

\bibitem[Pacini \& Salvati 1973]{ps73}
Pacini, F. \& Salvati, M. 1973, { ApJ}, {\rm 186}, 249.

\bibitem[Pivovaroff \etal  2000]{pkc+00}
Pivovaroff, M., Kaspi, V.~M., Camilo, F., Gaensler, B.~M., \& Crawford, F.
  2000, { ApJ}, {\rm }.
\newblock submitted.

\bibitem[Reynolds 1994]{rey94}
Reynolds, J.~E. 1994, { ATNF Technical Document Series}, {\rm 39.3040}.
\newblock
  (http://www.narrabri.atnf.csiro.au/observing/users\_guide/html/node215.html).

\bibitem[Reynolds \& Chevalier 1984]{rc84}
Reynolds, S.~P. \& Chevalier, R.~A. 1984, { ApJ}, {\rm 278}, 630 (RC84).

\bibitem[Sault \& Killeen 1999]{sk98}
Sault, R.~J. \& Killeen, N. E.~B. 1999, { The Miriad User's Guide}, (Sydney:
  Australia Telescope National Facility).
\newblock (http://www.atnf.csiro.au/computing/software/miriad/).

\bibitem[Seward, Harnden, \& Helfand 1984]{shh84}
Seward, F.~D., Harnden, F.~R., \& Helfand, D.~J. 1984, { ApJ}, {\rm 287}, L19.

\bibitem[Seward \& Harnden~Jr. 1982]{sh82}
Seward, F.~D. \& Harnden~Jr., F.~R. 1982, { ApJ}, {\rm 256}, L45.

\bibitem[Staelin \& Reifenstein 1968]{sr68}
Staelin, D.~H. \& Reifenstein, {III}, E.~C. 1968, { Science}, {\rm 162}, 1481.

\bibitem[Stappers, Gaensler, \& Johnston 1999]{sgj99}
Stappers, B.~W., Gaensler, B.~M., \& Johnston, S. 1999, { MNRAS}, {\rm 308},
  609.

\bibitem[Trimble 1968]{tri68}
Trimble, V. 1968, { AJ}, {\rm 73}, 535.

\bibitem[Whiteoak \& Green 1996]{wg96}
Whiteoak, J. B.~Z. \& Green, A.~J. 1996, { A\&AS}, {\rm 118}, 329.
\newblock (http://www.physics.usyd.edu.au/astrop/wg96cat/).

\bibitem[Zaninetti 2000]{zan00}
Zaninetti, L. 2000, { A\&A}, {\rm 356}, 1023.

\end{thebibliography}

%\clearpage

\begin{deluxetable}{lcccclccl}
\footnotesize
\tablecaption{
Pulsars with $\tau_{c} < 5$ kyr 
%(ranked by increasing characteristic age)
\label{chap6:tab1}}
\tablehead{
\colhead{Pulsar} & 
\colhead{$P$\tablenotemark{a}} &
\colhead{$\tau_{c}$\tablenotemark{b}} &
\colhead{$B$\tablenotemark{c}} &
\colhead{$\dot{E}$\tablenotemark{d}} &
\colhead{Associated SNR/PWN} &
\colhead{Radio} &
\colhead{Radio} &
\colhead{Reference} \nl
\colhead{} &
\colhead{(ms)} & 
\colhead{(kyr)} &
\colhead{(10$^{12}$ G)} &
\colhead{(10$^{36}$ erg s$^{-1}$)} &
\colhead{} &
\colhead{SNR?} & 
\colhead{PWN?} & 
\colhead{}
}
\startdata 
J1846--0258\tablenotemark{e}  & 324    & 0.7  &   48 &  8 & G29.7--0.3 (Kes~75) & Y & Y & Gotthelf \etal (2000\nocite{gvbt00}) \nl
B0531+21           & 33 &       1.3  &    3.8 & 450 & G184.6$-$5.8 (Crab
Nebula ) & N & Y  & Staelin \& Reifenstein (1968\nocite{sr68}) \nl
B1509--58            & 150 &       1.6  &  15 &  18 & G320.4$-$1.2 (MSH
15--5{\em 2}) & Y & N  & Seward \& Harnden (1982\nocite{sh82}) \nl
%J1119$-$6127 & 408 &  1.6  &  41 &   2 & G292.2$-$0.5        & Y & N  
%& {\bf Camilo et al.\ (2000a\nocite{ckl+00}); this paper} \nl
{\bf J1119$-$6127} & {\bf 408} &  {\bf 1.6}  &  {\bf 41} & 
{\bf 2} & {\bf G292.2$-$0.5}        & {\bf Y} & {\bf N} & {\bf Camilo et
al.\
(2000a\nocite{ckl+00}); this paper} \nl
B0540--69           &  50    & 1.7  &    5.0 & 150 & G279.7--31.5 (0540--69.3)         & Y & Y  & Seward, Harnden \& Helfand (1984\nocite{shh84}) \nl
J0537$-$6910\tablenotemark{e} & 16 & 5.0  &   0.9 & 480 & G279.6--31.7
(N~157B) & N & Y  & Marshall \etal (1998\nocite{mgz+98}) \nl % \hline
%B1610--50              & 231    & 7.4  & 11 &   2 & $\ldots$      & N & N  \nl
%J1617$-$5055             & 69     & 8.0  &  3.1 &  19 &  $\ldots$      & N & N  \nl 
%{\bf J1119$-$6127} & {\bf 408} &  {\bf 1.6}  &  {\bf 41} & 
%{\bf 2} & {\bf G292.2$-$0.5}        & {\bf Y} & {\bf N} & {\bf Camilo et
%(2000a\nocite{ckl+00}); this paper} \nl
\enddata 

\tablecomments{Entries ranked by characteristic age.}

\tablenotetext{a}{Period of rotation.}

\tablenotetext{b}{Characteristic age, $\tau_c \equiv P/2\dot{P}$.} 

\tablenotetext{c}{Inferred surface dipolar magnetic field, $B \approx 
3.2\times 10^{19}~(P\dot{P})^{1/2}$~G.}

\tablenotetext{d}{Spin-down luminosity, $\dot{E} \approx 4\pi^2 \times 
10^{45}~\dot{P}/P^3$~erg~s$^{-1}$.}

\tablenotetext{e}{X-ray pulsar only.}

\end{deluxetable}

\begin{deluxetable}{lcccc}
\footnotesize
\tablecaption{ATCA observations of PSR J1119$-$6127\label{chap6:tab2}}
%\tablewidth{0pt}
\tablehead{
\colhead{Observing} &
\colhead{Observing Frequencies} &
\colhead{Array} & 
\colhead{Time on Source} \nl
\colhead{Dates} &
\colhead{(MHz)} &
\colhead{Configuration} & 
\colhead{(h)} 
}
\startdata
1998 Oct 30-31 & 1384/2496  & 6D     & 9 \nl
1998 Nov 1     & 1344/2240  & 6D     & 5  \nl 
1999 Nov 25-26 & 1384/2496  & 0.375  & 9 \nl
1999 Dec 14    & 1384/2496  & 1.5A   & 9 \nl
\enddata

\end{deluxetable}

\begin{deluxetable}{llcccc}
\footnotesize
\tablecaption{Sizes and surface brightnesses of selected
radio pulsar wind nebulae\label{chap6:tab3}}
\tablehead{
\colhead{PWN or} & 
\colhead{Other} &
\colhead{$d$\tablenotemark{a}} &
\colhead{$\theta$\tablenotemark{b}} &
\colhead{$S_{\rm 1~GHz}$\tablenotemark{c}} &
\colhead{$\log \Sigma$\tablenotemark{d}} \nl
\colhead{Associated SNR} & 
\colhead{Name} &
\colhead{(kpc)} & 
\colhead{(arcmin)} &
\colhead{(Jy)} & 
\colhead{(W~m$^{-2}$~Hz$^{-1}$~sr$^{-1}$)} 
}
\startdata 
G184.6$-$5.8                 & Crab Nebula, SN~1054 &            2.5 & 6                     & 1040 & $-$17.4                    \nl 
G279.7--31.5 & 0540$-$69.3                & 50  & 0.08                 & 0.09 & $-$17.7                    \nl

G21.5$-$0.9                  &                   & 5.5 & 1.2                   & 6    & $-$18.2                    \nl 
G5.4$-$1.2\tablenotemark{e}  &               & 5   & 0.07                 & 0.01 & $-$18.5                    \nl  
G279.6--31.7      & N~157B             & 50  & 1.5                   & 2.75 & $-$18.7                    \nl
G29.7--0.3                   & Kes~75             & 19 & 0.5 & 0.3 & --18.8 \nl
G328.4$+$0.2                 & MSH 15$-$5{\em 7} & 17  & 5                     & 16   & $-$19.0                    \nl
G130.7$+$3.1                 & 3C~58              & 3 & 8                     & 33   & $-$19.1                    \nl
G34.7$-$0.4\tablenotemark{e} & W~44               & 3   & 1                     & 0.3  & $-$19.4                    \nl
G326.3$-$1.8                 & MSH 15$-$5{\em 6} & 4   & 7                     & 14   & $-$19.4                    \nl 
G74.9$+$1.2                  & CTB~87             & 12  & 7                     & 9    & $-$19.6                    \nl
G320.4$-$1.2\tablenotemark{f}                 & MSH 15$-$5{\em 2} & 5 & 4 &  $\ldots$    & $< -19.6$ \nl
G292.2$-$0.5\tablenotemark{f}                 &                   & $\sim$5   & 4 &  $\ldots$    & $-$22.0  \nl     
\enddata 

\tablecomments{Data from Green~(2000)\protect\nocite{gre00} and references therein.} 

\tablenotetext{a}{Distance to PWN.}
\tablenotetext{b}{Mean angular diameter of PWN.}
\tablenotetext{c}{1-GHz flux density of PWN.} 

\tablenotetext{d}{1-GHz surface brightness of PWN.}

\tablenotetext{e}{Bow-shock nebula.} 

\tablenotetext{f}{Radio PWN not detected. Angular size
is derived using expansion velocity of Crab Nebula
($V \sim 1700$ km/s) and characteristic age of associated pulsar.
For G320.4--1.2, 
upper limit on surface brightness is that 
obtained by Gaensler \etal (1999\nocite{gbm+98});
for G292.2--0.5, surface brightness is that
predicted by the model described in the text.}

\end{deluxetable}

\begin{deluxetable}{lll}
\footnotesize
\tablecaption{Input parameters to PWN model
\label{chap6:tab4}}
\tablehead{ 
\colhead{Parameter} & 
\colhead{PSR~B0531+21 / Crab Nebula} &
\colhead{PSR J1119$-$6127} 
} 
\startdata 
Initial pulsar spin period, $P_{0}$ (ms) & 16 & 16 (assumed)          \nl
Initial surface magnetic field, $B_{0}$ (G) & 3.8 $\times 10^{12}$ & 4.1 $\times 10^{13}$  \nl
Pulsar age, $t$ (kyr)                         & 1.0  & 1.7        \nl 
Pulsar braking index, $n$  & 2.51 & 2.91 \nl
PWN expansion velocity, $V$ (\kms)             & 1700    & 1700 (assumed)            \nl 
\enddata 

\end{deluxetable}

\clearpage

\end{document}